\documentclass{wlscirep}
\usepackage[utf8]{inputenc}
\usepackage[T1]{fontenc}
\title{A reconfigurable non-linear active metasurface for coherent wave down-conversion\\[0.2in]}

\author[1]{Pouria Sanjari}
\author[1,*]{Firooz Aflatouni}

\affil[1]{Department of Electrical and Systems Engineering, University of Pennsylvania, Philadelphia, PA 19104, USA}

\affil[*]{firooz@seas.upenn.edu}

\DeclareCaptionLabelSeparator{pipe}{ $|$ }
\begin{abstract}
\textbf {\small Metasurfaces can manipulate the amplitude and phase of electromagnetic waves, offering applications ranging from antenna design and cloaking to imaging and communication. Additionally, temporal, and non-linear metasurfaces have the potential to adjust the frequency of impinging waves, driving advancements in frequency conversion, sensing, and quantum systems. Here, we report the demonstration of a non-linear active electronic-photonic metasurface that transfers information from an impinging optical wave to a millimeter-wave (mm-wave) beam. The proof-of-concept metasurface is designed to radiate a steerable 28GHz beam when illuminated with an optical wave at 193THz and consists of optically synchronized electronic-photonic chips tiled on a printed circuit board containing a microstrip patch antenna array. Input light, modulated with a data-encoded mm-wave carrier, is coupled into electronic-photonic chips using microlenses. Within each chip, the mm-wave signal is detected, phase-adjusted, amplified, and routed to an off-chip antenna. Beam-steering over a range of 60$^\circ$ in elevation and azimuth and data transmission at 2Gb/s over a fiber-wireless link is demonstrated. Free-space optical synchronization can significantly reduce the complexity of large-scale metasurfaces composed of non-uniform or randomly placed elements, is compatible with scalable architectures, and facilitates data transfer and mm-wave beam shaping, allowing for large-scale high-bandwidth and energy-efficient links with reduced complexity for the next generation communication, computation, sensing and quantum systems.
} 
\end{abstract}

\begin{document}

% Pouria - overwrite settings
\def\figurename{Fig.}
\captionsetup{justification=justified} 
\captionsetup{labelsep=pipe}

\flushbottom
\maketitle
\thispagestyle{empty}

\section*{Introduction}

\noindent Metasurfaces are two-dimensional planar structures with subwavelength constituent components that have been arranged periodically, quasi-periodically, or even randomly\cite{kildishev2013planar}. These engineered structures have often been used to manipulate electromagnetic (EM) waves in compelling ways, consequently facilitating innovations in optics\cite{zheng2015metasurface,huang2013three,arbabi2018mems,yin2017beam}, communication\cite{zhao2020metasurface,zhang2021wireless,shlezinger2021dynamic}, computation\cite{wu2021programmable,mohammadi2019inverse,zangeneh2021analogue}, and various other diverse fields\cite{cai2007optical,alu2005achieving,hawkes2013microwave}. Compared to their bulk counterparts, metasurfaces are much thinner, easier to fabricate, and highly scalable, enabling compact integration for photonic and electronic systems.\par
The main utility of metasurfaces stems from their ability to control various properties of impinging EM waves including amplitude, phase, polarization, and momentum\cite{kildishev2013planar,yu2011light}. Moreover, these structures can be designed to exhibit strong nonlinearity either through field enhancement and localization effects, predominantly observed in the optical domain\cite{glybovski2016metasurfaces}, or by embedding a non-linear element such as a varactor\cite{luo2019intensity,rose2011controlling} which is commonly used in the microwave regime. This versatility makes metasurfaces appealing for diverse applications, spanning from frequency conversion and harmonic generation\cite{liu2016resonantly} to non-linear imaging, holography\cite{zheng2015metasurface,huang2018metasurface}, non-linear sensing\cite{qin2022metasurface}, and quantum systems\cite{wang2018quantum,stav2018quantum}.\par
In programmable and reconfigurable metasurfaces\cite{liu2018programmable, oliveri2015reconfigurable}, the surface properties can be actively tuned allowing for real-time adjustments of the incident EM wave properties and control over the radiation pattern of the scattered wave. This capability has sparked new paradigms in sensing\cite{dabidian2016experimental,lin2021single}, computation\cite{wu2021programmable}, computational imaging\cite{watts2014terahertz,li2016transmission}, and adaptive optics\cite{ee2016tunable}. RF wireless communication is another promising avenue for such structures\cite{tang2020wireless}. To date, numerous technologies with functionalities such as multi-beam communication\cite{zhang2021wireless} and beam scanning\cite{cui2014coding}, which are frequently discussed for next-generation wireless systems, have been devised and demonstrated all while benefitting from simplified architectures, lower hardware costs, and reduced complexities.\par
Like metasurfaces, phased array antenna systems can be used to shape the wavefront of electromagnetic waves and wirelessly transmit data. While the radiated beam in metasurfaces is formed through manipulation of an input impinging electromagnetic waves, phased array antennas require the distribution of a reference signal to time/phase synchronize the array elements in order to form a coherent beam. For such systems, element synchronization becomes more difficult due to the challenges of routing high-frequency electrical interconnects at the package and board level. Crosstalk, EM interference, ohmic and substrate losses, and increased power consumption (due to the required buffering) are some of the issues that can contribute to performance deterioration. With the ever-increasing data rate and speed requirements of advanced and emerging communication systems, it is thus more pertinent to design and develop scalable metasurfaces that can operate at high frequencies while also offering reduced complexities and enhanced power efficiencies.\par

On the front of pioneering solutions and concepts for the ubiquitous realization of efficient high-data-rate communication systems, and more recently computation systems as well, the adoption and confluence of both optical and electrical technologies has consistently been regarded as integral\cite{capmany2007microwave,ashtiani2022chip,shastri2021photonics}. Coupled with the versatility of electronics, photonics may be employed to assist in the generation, distribution, and processing of signals\cite{yao2009microwave,marpaung2019integrated,capmany2012microwave}. The large instantaneous bandwidth available around an optical carrier, the immunity to EM interference, and the availability of low-loss optical transmission mediums can all be leveraged to aid system implementation and or attain performance metrics that are inconceivable with traditional all-electronic circuitry. To date, numerous designs with all-optical delay and phase control based on switched fiber delay lines\cite{yu2014multi}, optical dispersion in fiber\cite{akiba2014photonic} and fiber gratings\cite{zhang2016photonic}, and photonic microwave phase shifters\cite{yi2011photonic} have been reported demonstrating photonic beamforming with wide operational bandwidths and enabling capabilities such as remotely-fed fiber-distributed phased array antennas. Though despite their excellent performance, the adoption of these technologies has often been limited given their reliance on discrete, power-hungry, and bulky components. More recently, integrated designs based on switched waveguides\cite{zhu2020silicon,zheng2019seven}, ring resonators\cite{zhuang2010novel} and Mach–Zehnder delay interferometers\cite{duarte2016photonic} aimed at realizing true-time delay lines (TTDL) for photonic beamformation have been devised. Such systems have great potential for scalability and high-yields and moreover can mitigate the issue of beam squint, particularly desirable for broadband phased arrays, but often result in increased complexities and chip area given the use of TTDL elements.  Advantageously, for large-span array, long haul applications and or radio-over fiber (RoF) systems\cite{novak2015radio}, the previously described optically enabled schemes may directly interface with existing optical fiber infrastructure. This allows for the distribution of high frequency signals from a central station to a remote antenna unit for remote synchronization\cite{gal2022optically}, beamformation\cite{akiba2014photonic} and wireless radiation, essential to support cost-effective, high-capacity links and provide high-speed interconnectivity. In light of these advantages, advancements in hybrid optoelectronic metasurfaces, aimed at realizing hybrid transmitters for wireless communication, have also recently emerged\cite{zhang2022metasurface,zhang2020optically}. While these advancements demonstrate significant progress and utility for low-cost and low-complexity solutions, they still fall short in achieving the targeted performance required by multi-Gb/s links compared to, for instance, all-electronic millimeter-wave (mm-wave) phased arrays\cite{kibaroglu201864}. 

In this article, we report the demonstration of a reconfigurable non-linear active metasurface that can radiate a data-carrying, steerable mm-wave beam when illuminated with an optical wave that has been modulated with the data encoded mm-wave signal. More specifically, the metasurface is designed to radiate a mm-wave beam at 28GHz when incident with a modulated optical wave operating around 193THz and is comprised of optically synchronized electronic-photonic chips and a printed circuit board antenna array. To our knowledge, this is the first highly scalable, reconfigurable, low-energy, hybrid non-linear metasurface with demonstrated applications in wavefront front shaping and multi-Gb/s fiber-wireless communication. Free-space optical synchronization of the surface elements eliminates the electrical interconnect network, often utilized in all-electrical phased array antennas, leading to reduction in the overall complexity of the system, size, power consumption and undesired EM coupling and interference. Moreover, it enables the practical implementation of large scale metasurfaces with sparsely or non-uniformly placed elements. Compared to metasurfaces and arrays where the comprising elements are synchronized using fiber interconnects, free-space offers greater flexibility in cases where physical routing may be impractical or undesirable. For such systems with many elements, packaging challenges may also arise when the fiber must be uniformly split and separate cores with equal lengths must be aligned with the various constituent elements. Additionally, fibers aligned and or glued are more fragile and sensitive to tights bends, leading to radiation losses, especially in compact form factors\cite{goodman1984optical}. The implemented proof-of-concept metasurface consists of four electronic-photonic integrated circuit (EPIC) chips, each incorporating a photonic receiver and mm-wave transmitter, fabricated using GlobalFoundries 90nm CMOS silicon photonic process, and a low-cost FR4 printed circuit board with a microstrip patch antenna array. The challenge with free-space synchronization often stems from the fraction of optical power that is converted by the comprising elements. To address this, microlenses are used to focus the impinging light on the grating couplers of the EPICs and enhance the optical coupling into the chips. Photodetectors, integrated on the chips, convert the modulated optical wave - carrying both a millimeter-wave carrier and data - into an electrical current through their second-order (square-law) non-linear operation on the optical electric fields. The recovered data-carrying millimeter-wave signals are subsequently amplified and phase-adjusted by the RF transmitter chains of the EPICs (operating in the linear regime) before being radiated by the antennas. Following this process, by controlling the relative phase of adjacent EPIC elements, millimeter-wave beams that can be actively steered in both elevation and azimuth can be formed in the far-field of the antenna array. Using the implemented metasurface, we demonstrate beam steering with a scanning range of around 60${^\circ}$ in elevation and azimuth and a fiber to wireless communication link where high-order modulation signal (32-QAM) is transmitted over fiber and wirelessly radiated at 28GHz in various directions achieving data rates as high as 2Gb/s. The presented non-linear reconfigurable active metasurface, which is comprised of optically synchronized electronic-photonic elements, facilitates data transfer and mm-wave beam shaping while being compatible with highly scalable architectures and has the potential to realize energy-efficient, compact links with reduced complexity for the next generation communication, computation, sensing and quantum systems. 

%%%%%%%%%%%FIGURE%%%%%%%%%%%%%%%%%%
\begin{figure}[!t]
\begin{center}
\includegraphics{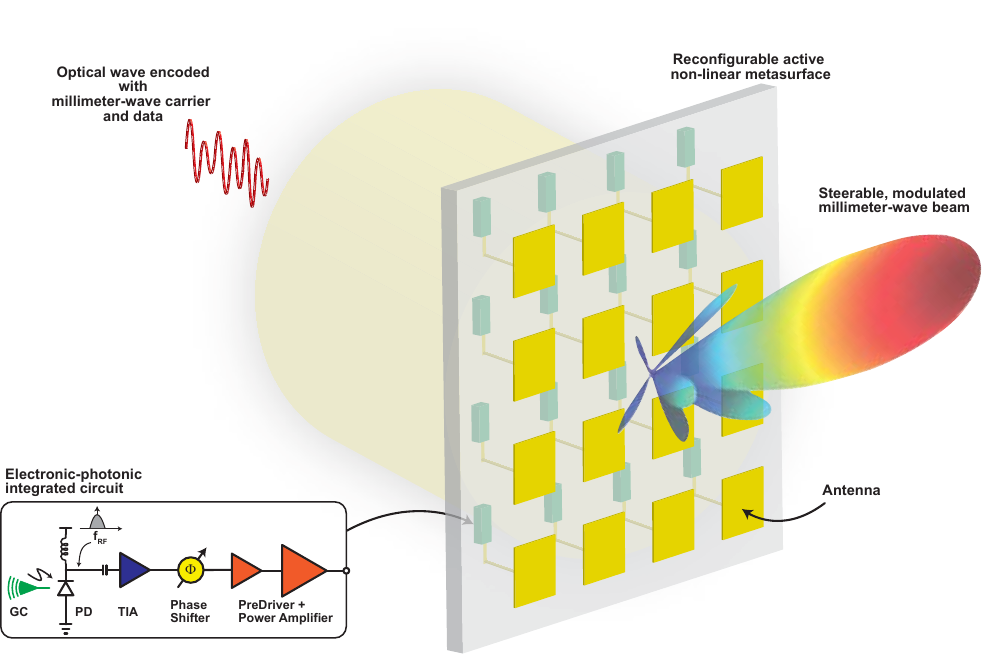}
\caption{\textbf{General concept of the reconfigurable non-linear active metasurface.} The metasurface is comprised of electronic-photonic chips and antenna elements. When illuminated with an optical wave that has been encoded with mm-wave carrier and data signals, the metasurface radiates a data-carrying mm-wave beam which can be scanned in space (GC: grating coupler; PD: photodetector; TIA: trans-impedance amplifier).} \label{fig:1}
\end{center}
\end{figure}
%%%%%%%%%%%%%%%%%%%%%%%%%%%%%

\section*{Results}
\subsection*{A hybrid non-linear metasurface}
The general concept of the reconfigurable non-linear active metasurface is illustrated in Fig. 1. The metasurface consists of a low-cost printed circuit board (PCB) with electronic-photonic chips tiled on the front side and a microstrip patch antenna array printed on the backside. Light from a laser operating at 1550nm, which has been modulated with a mm-wave carrier encoded with data, impinges on the EPICs. The modulated optical wave couples into the chips through grating couplers, is guided by silicon nanophotonic waveguides, and is photodetected by on-chip photodetectors, recovering the mm-wave carrier and data signals. Besides the front-end integrated photonic receiver, the EPIC also contains a single-channel mm-wave transmit chain, consisting of amplifiers and a phase shifter, that is used to boost and control the phase of the photodetected signal. The output of an EPIC is connected to the feedline of a patch antenna which radiates the data-carrying mm-wave signal. Given that the EPICs receive the same incident light, the mm-wave signals recovered by the photodetectors of the different chips are inherently coherent, and thus the various transmitter elements are synchronized. As such, by controlling the relative phase between adjacent elements, a mm-wave beam that can be actively steered is formed in the far-field of the antenna array. With the presented scheme, all high-frequency routing interconnects, and chip-to-board transitions, except for the chip-to-antenna feedline and antenna itself, are effectively eliminated. This greatly simplifies the design process of the EPIC and antenna array, circumvents the challenges of on-PCB mm-wave routing such as EM interference and unwanted coupling, and can facilitate the implementation of metasurfaces with non-uniform, sparse or randomly placed elements. Furthermore, the board substrate and stackup can be independently optimized for the antenna performance adding extra flexibility as the only other electrical interconnects required are for low-frequency control signals and DC supply lines. Moreover, with the recent advancements of CMOS-silicon photonics technology, photodetectors with electro-optic bandwidths exceeding 50GHz and high-performance electronic technology nodes can now be monolithically integrated on the same platform. This in turn allows high-frequency mm-wave signals to be directly distributed on-chip (for phase adjustment and amplification) eliminating the need for power-hungry and interference-prone mm-wave frequency multipliers and/or phase-locked loops that can limit the aggregate performance and increase chip area. Finally, as another distinct advantage, the proposed scheme is compatible with the previously discussed RoF networks\cite{novak2015radio}. That is, an optical fiber can be used to transfer mm-wave and high-data-rate bitstreams from a centralized location - housing all the complex and power-hungry signal generation and processing tasks – to interface directly with a remote, low-cost, low-complexity metasurface that simultaneously leverages the optical carrier to synchronize its elements thus enabling wireless beamformation and active scanning.

%%%%%%%%%%%FIGURE%%%%%%%%%%%%%%%%%%
\begin{figure}[!t]
\begin{center}
\includegraphics[width=176mm]{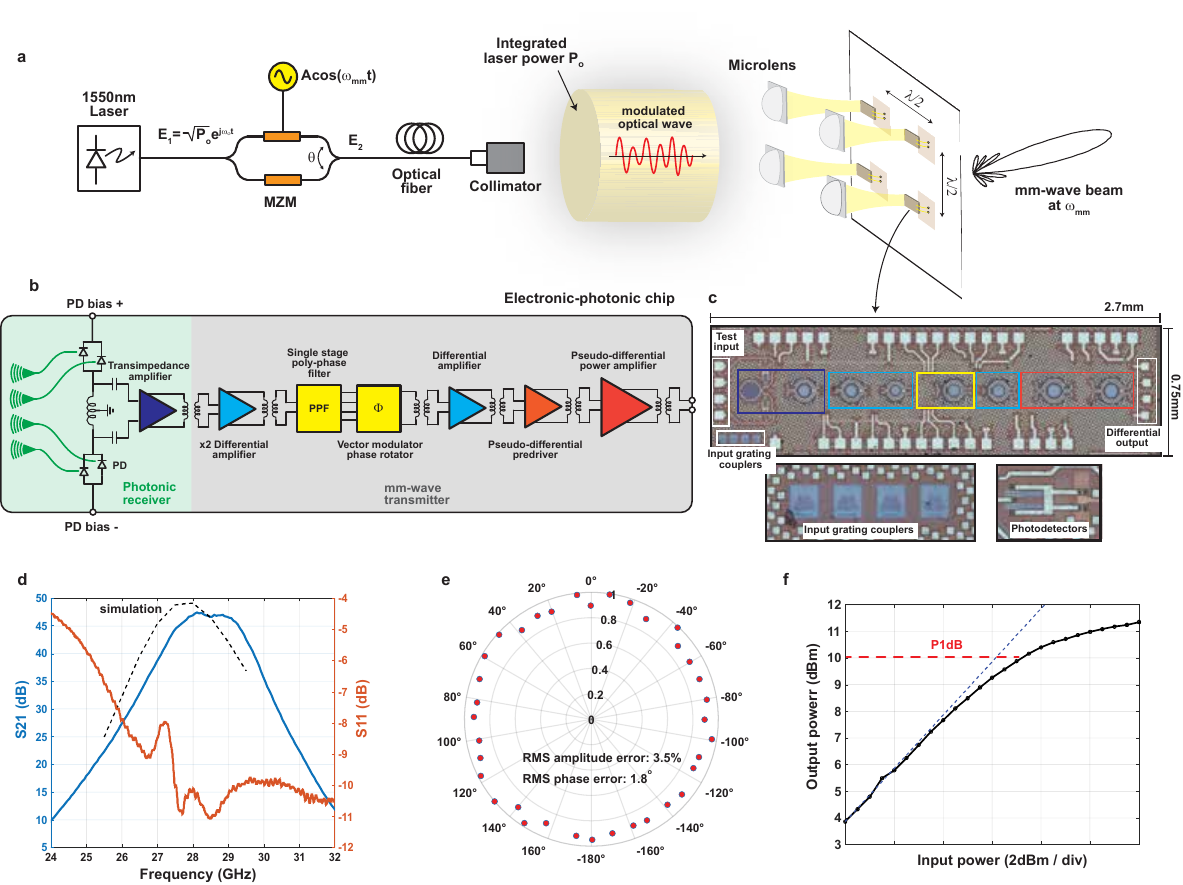}
\caption{\textbf{System level architecture and characterization of the electronic-photonic chip.} \textbf{a.} Illustration of a setup where a Mach-Zehnder modulator (MZM) intensity modulates light from a continuous-wave laser encoding a mm-wave signal at $\omega_{mm}$ on the optical wave. The modulated optical wave is transmitted through an optical fiber to the metasurface, where a collimator expands and collimates the light creating an optical beam that illuminates the electronic-photonic chips. The chips have been tiled on the topside of the PCB and wire bonded to the feedlines of patch antennas which are printed on the bottom side and spaced 0.5$\lambda$ ($\lambda$ corresponds to the wavelength of the modulating mm-wave signal). Microlenses focus the light on grating couplers of the electronic-photonic chips. If the phase difference $\theta$ at the output arms of the MZM is set to $\pi /2$, the photodetected current at the output of the on-chip photodetectors will have an AC component at $\omega_{mm}$ \textbf{b.} Block diagram of the electronic-photonic chip consisting of the photonic receiver and mm-wave transmitter. \textbf{c.} Micrograph of the electronic-photonic chip monolithically fabricated on GlobalFoundries 9WG process. \textbf{d.} S-parameter measurements of the chip (excluding trans-impedance gain) measured using the test input traces. \textbf{e.} Vector modulator phase shifter performance after calibration. \textbf{f.} Power amplifier output power at 28GHz.} \label{fig:2}
\end{center}
\end{figure}
%%%%%%%%%%%%%%%%%%%%%%%%%%%%%

\subsection*{System implementation}
The system diagram of the proposed metasurface employing optics, photonics and electronics is shown in Fig. 2a. Here, the output of a laser is intensity-modulated by a mm-wave signal using an electro-optic Mach-Zehnder modulator (MZM)\cite{thomaschewski2022pockels}. The up-converted mm-wave signal is transmitted down on an optical fiber to the metasurface where the modulated light is expanded and collimated to a spot size large enough to illuminate the EPIC elements that have been tiled on the top side of the printed circuit board. The impinging light is coupled into each chip through grating couplers. To enhance the coupling efficiency from free-space into the chips, microlenses are used. In this scenario, as illustrated in Fig. 2a, each lens collects and focuses a larger amount of power on the grating couplers. Since the ratio of the lens area to the beam area is much larger, and the focused spot size can be smaller than the size of the grating coupler, the optical power coupled into the chip significantly increases. The modulated optical wave that is coupled to each EPIC chip is guided to photodetectors with silicon nanophotonic waveguides. 
The photocurrent generated at the output of a photodetector has a mm-wave component which is amplified and converted to voltage using a trans-impedance amplifier (TIA) with a tuned response around the mm-wave frequency. Within a chip, gain stages further amplify the TIA output, a phase shifter controls the phase of the mm-wave signal, and a power amplifier (PA) drives an antenna that has been printed on the bottom side of the PCB and placed on a $\lambda /2$ square grid to avoid grating lobes where $\lambda$ corresponds to the wavelength of the mm-wave signal. Since the same modulated light impinges on all the chips, the photodetected mm-wave signals are coherent and phase-synchronized. Consequently, by controlling the relative phase between the antenna elements, which is done electronically by adjusting the phase shifter settings, a mm-wave beam that can be steered in 2D (elevation and azimuth) is formed in the far-field of the antenna array. Further details of the mm-wave detection and microlens alignment are presented in the Methods sections.\par
In the above implementation, given the small ratio of the grating coupler area to the collimated beam spot size, the resulting coupling loss per chip is around 58.5dB (including the loss of the grating coupler) and hence the amount of optical power coupled from free-space into the chip is quite small. Through the manual alignment of the microlenses, we were able to increase the amount of optical power coupled to a chip by around 19dB. Since the power of the detected mm-wave signal is proportional to the square of the coupled optical power, the equivalent isotropic radiated power (EIRP) increases by around 38dB. Note that although the microlenses enhance the optical coupling by orders of magnitude, further improvements in the coupling efficiency could still be expected if instead of the beam expander-collimator, a diffractive optical element is used. In this scenario, the diffractive optical element can be used to create spot arrays where the diffraction pattern, illuminating the grating couplers, can contain most of the incident power\cite{katz2018using}.

\subsection*{Electronic-photonic integrated chip}
The electronic-photonic integrated circuit used in the metasurface was designed and monolithically integrated in GlobalFoundries Silicon Photonics 90nm SOI CMOS process. Figure 2b shows the architecture of the EPIC that consists of a photonic receiver followed by a mm-wave transmitter. Differential design of the various electronic blocks is preferred to ensure circuit stability and reduce the parasitic effects from ground and supply bondwires. Additionally, transformer-based matching networks are used to couple the various stages of the transmitter chain.\par
The modulated optical wave is coupled into the photonic receiver through a grating coupler, optimized for TE-polarization, with a coupling loss of around 5.5dB, and is detected using a SiGe p-i-n photodetector with a responsivity of around 1A/W. Note that in the absence of a microlens (which focuses light on one of the grating couplers of an EPIC), the collimated optical beam illuminates all four grating couplers of an EPIC chip (two on each side of a differential input). In this case, the extra grating coupler on each side effectively increases the optical power coupled to the chip, which is useful for initial alignment and optimization of the collimator angle, and the photocurrents of the two photodetectors (which are in parallel) on each side are added in the current domain. With this configuration, as shown in Fig. 2b, bipolar biasing of the photodetectors may also be used to generate a differential signal at the output of the photodetectors\cite{li2020co}. The mm-wave component of the generated photocurrent is amplified and converted to a voltage using a trans-impedance amplifier (TIA). In the input of the photonic receiver, a symmetric inductor with a grounded center tap reverse biases the photodetectors and resonates with the capacitances of the photodetectors and input of the amplifier at 28GHz. The TIA output voltage is then routed to the mm-wave transmitter, where it is amplified, and phase adjusted.\par
Note that while the chip was designed to be operated with an optical input, it could be characterized through an auxiliary electrical input. This differential test input bypasses the photodetectors and connects to input of the first stage amplifier. Figure 2d shows the S-parameter measurements of the chip indicating a gain of around 47dB (excluding the trans-impedance gain) with a bandwidth of approximately 1.5GHz around 28.5 GHz. Specific details of the characterization setup including the effect of the test input are further discussed in the Methods section.\par
Two cascaded variable gain amplifiers (VGA) form the input of the mm-wave transmitter with the second used as a buffer for the phase shifter. Phase adjustment (and thus beam steering) is achieved electronically using a vector modulator (VM) phase shifter. The phase shifter has a measured root-mean-square (RMS) phase error of around 1.8$^{\circ}$ and an average RMS amplitude error of 3.5\% after calibration (Fig. 2e). A differential gain stage and a pseudo-differential pre-driver and power amplifier follow the phase shifter and are used to further amplify the output of the phase shifter and drive the off-chip antenna. The power amplifier is designed for linear operation and has a measured output 1dB compression point of around 10dBm (Fig. 2f).\par 
The EPIC was monolithically integrated within a footprint of 2mm$^2$ and has an average power consumption of around 177mW. The details of the various circuit blocks, including circuit topology and characterization results, are provided in the Methods section and Extended Data Fig. 1.

%%%%%%%%%%%FIGURE%%%%%%%%%%%%%%%%%%
\begin{figure}[!t]
\begin{center}
\includegraphics[width=176mm]{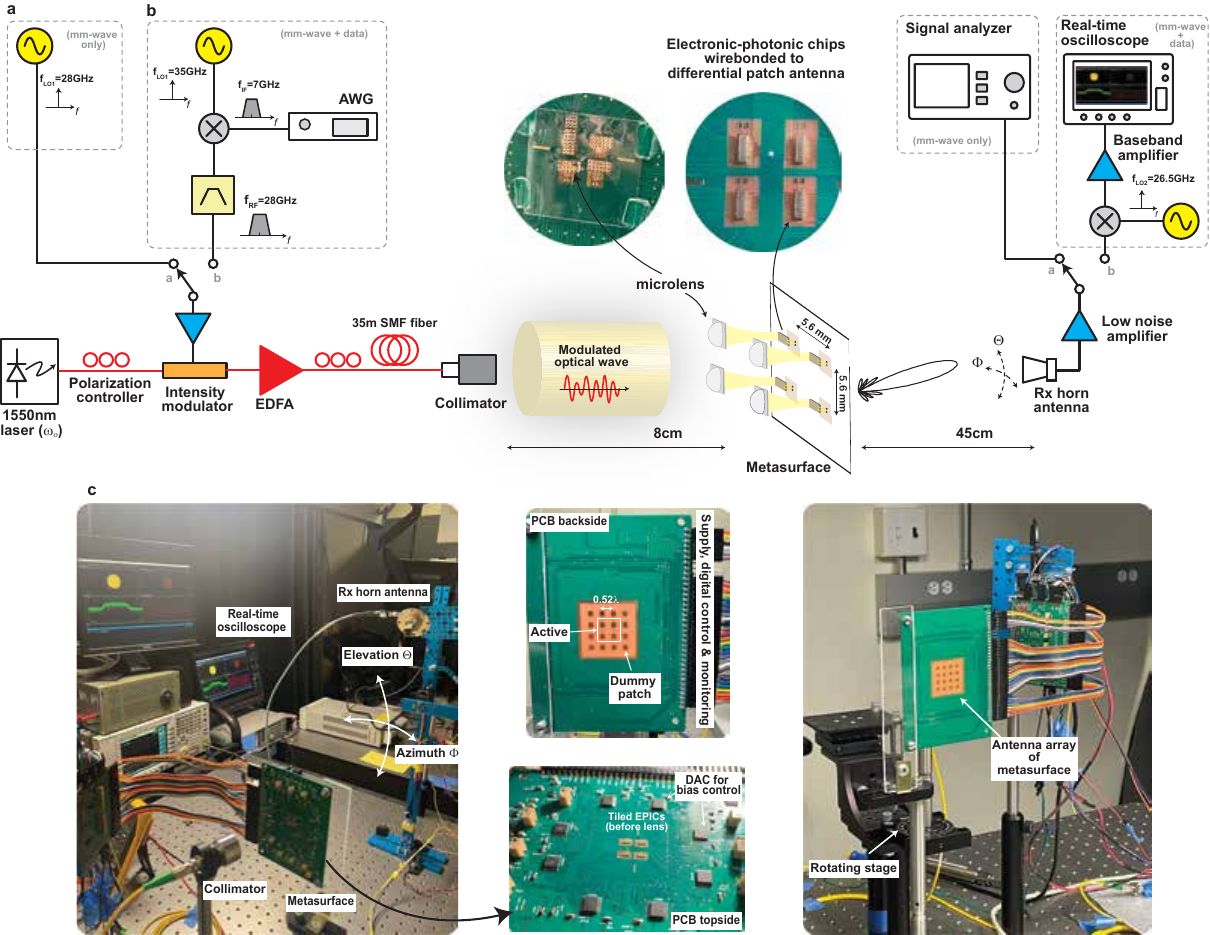}
\caption{\textbf{Measurement setup for beamformation and data communication.} \textbf{a.} Illustration of the setup used to measure the radiation pattern and demonstrate data communication over a fiber-wireless link using the implemented metasurface. The horn antenna has been mounted on a robotic arm and can scan the surface of a half-sphere to receive the radiated mm-wave signal at different elevation and azimuth angles. \textbf{a.} The first configuration (switch in position a) is used for radiation pattern and EIRP measurements given that only the mm-wave carrier signal modulates the laser light. \textbf{b.} The second configuration (switch in position b) is used to demonstrate a communication link where a QAM signal is generated using the Arbitrary Waveform Generator (AWG), up-converted and used to modulate the laser light. A receive chain consisting of a down-converting mixer, baseband amplifier, and real-time oscilloscope is used to demodulate the signal and capture the constellation. \textbf{c.} Images of the measurement setup (DAC: Digital to analog converter).} \label{fig:3}
\end{center}
\end{figure}
%%%%%%%%%%%%%%%%%%%%%%%%%%%%%

\subsection*{Beamformation and wireless data transfer}
The measurement setup displayed in Fig. 3 was used to demonstrate the functionality of the proof-of-concept reconfigurable non-linear active metasurface and its capability for beamformation and fiber to wireless data transfer. Here, a laser operating around 1550nm was intensity modulated with a 28GHz signal and amplified using an erbium-doped fiber amplifier (EDFA). The light at the EDFA output was then expanded and collimated. The collimated output beam has a diameter of around 7mm, large enough to illuminate four electronic-photonic chips that were placed 5.6mm apart corresponding to $\sim \lambda/2$ at 28GHz. Four silicon microlenses focused the collimated optical beam on the grating couplers of the four chips. The collimator was placed on a rotating stage to match the incidence angle of the grating couplers. For a low-cost solution, it is often desired for the chips to be tiled on a PCB with embedded antennas. Given that the length of the antenna feedlines can be very short, (since the ICs are placed close to the antenna), a low-cost FR4 substrate can be chosen to implement the differential patch antennas. The outputs of the chips (PA outputs) were then wire bonded to the feedlines which connect with vias to the differential patch antennas designed on the bottom side of a four-layer PCB with the inner layers used as ground planes. The differential design avoids any lossy balanced-unbalanced (balun) circuit structures at the output of the chip and simplifies the design of the differential transition vias from top layer to the bottom layer of the PCB. The four-element antenna array has a simulated gain of approximately 9dB, with the design details and simulation results presented in the Methods section and depicted in Extended Data Fig.2. To receive the radiated 28GHz signal, a horn antenna with a 10dB gain was placed 45cm away in the far-field of the array. The horn antenna was mounted on a robotic arm that could scan the surface of a half-sphere to measure the radiation pattern at different elevation and azimuth angles. The signal received by the horn antenna was amplified using a low noise amplifier and monitored on a signal analyzer. To form a beam and demonstrate steering, the electronic-photonic chips were turned on and the on-chip photodetectors were reversed biased at 1V. The polarization controller and the angle of the collimator were adjusted to maximize the optical coupling into the chip through the grating couplers. By adjusting the relative phase between the chips using the VMs, a mm-wave beam could be formed and scanned in space. The beam-steering measurement results are shown in Fig. 4a and Fig. 4b where about 60$^{\circ}$ scanning range is demonstrated in elevation and azimuth. Note that the steering range is effectively limited by the radiation pattern of a single patch antenna shown in Extended Data Fig. 2c. In this implementation, a higher gain was preferred at the cost of a narrower beamwidth. In this implementation, a higher gain was preferred and achieved by employing a thicker substrate, but at the cost of a narrower beamwidth. If desired, the radiation pattern can be broadened by modifying the patch antenna design\cite{ning2008microstrip} or by using, for example, parasitic elements\cite{su202079}. 
To further increase the gain and radiation efficiency of the patch antenna given the losses of the FR4 substrate, implementing cavities comprised of tightly spaced vias\cite{hwang2019cavity} or air structures\cite{kim2021design} beneath the radiating patch, or techniques utilizing spatial filters\cite{kim20194} and artificial magnetic surfaces\cite{zhang2003planar} may also be employed.\par
The equivalent isotropic radiated power (EIRP) of the metasurface was also calculated by measuring the received power, which was around 10dBm when operating with 200mW of optical power (Extended Data Fig. 4b). Details of EIRP calculations along with an estimate of the maximum wireless transmission distance are presented in the Methods section. \par
To demonstrate data transfer from input optical beam to the radiated mm-wave beam using the implemented metasurface, a modulated mm-wave (instead of continuous-wave) signal can be used to drive the intensity modulator. Figure 3 also depicts the measurement setup used to demonstrate the link using the implemented metasurface. In this scenario, an arbitrary waveform generator (AWG) was programmed to generate an M-ary waveform at an intermediate frequency (IF) of 7GHz. The IF signal was then upconverted to 28GHz using an external mixer and amplified before driving the optical intensity modulator. As before, the modulated optical signal was collimated and used to illuminate the metasurface. The data-carrying 28GHz mm-wave beam was radiated by the metasurface at different scan angles, received by the horn antenna and down-converted to an IF of 1.5GHz. Carrier-recovery and digital demodulation were then performed using a Keysight real-time oscilloscope running the Vector Signal Analysis (VSA) software. Figure 4c shows constellations and error-vector magnitudes (EVMs) at various scan angles when operating at 2Gb/s (400Mbaud) with a 32QAM waveform. Note that in these measurements, no signal pre-distortion at the AWG or equalization at the oscilloscope were employed. Detailed description of the radiation pattern measurement and communication setup are provided in the Methods section. Moreover, as elaborated in the Methods section (eq.\ref{eq11}), the signal-to-noise ratio (SNR) at the input of the trans-impedance amplifier was limited to around 17.5dB due to the low photodetected optical power. It is expected that by utilizing better equipment and enhancing optical coupling efficiency, the SNR can be increased, which can be used to support higher data-rates and improve system throughput.

%%%%%%%%%%%FIGURE%%%%%%%%%%%%%%%%%%
\begin{figure}[!t]
\begin{center}
\includegraphics[width=176mm]{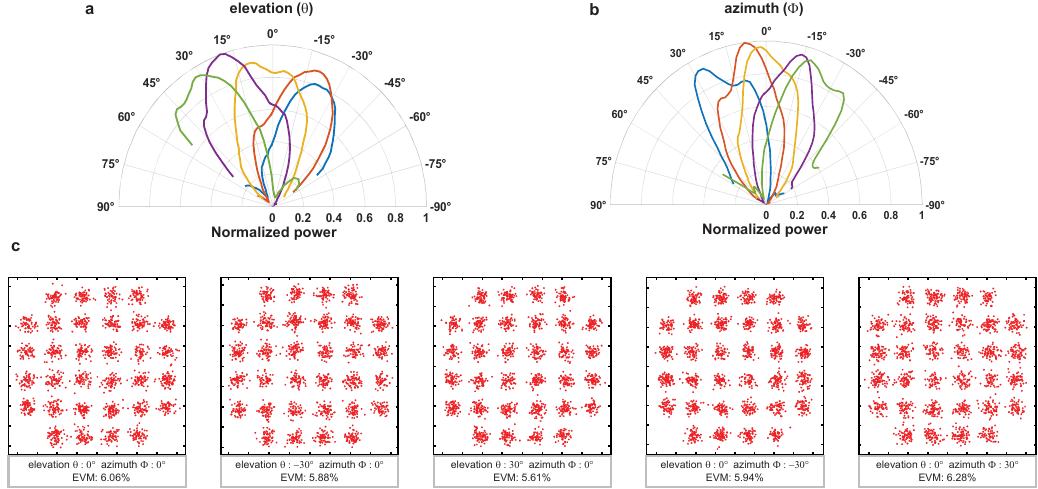}
\caption{\textbf{Measurement results of beam scanning and data communication.} Measured radiation patterns in \textbf{a.} elevation ($\theta$) and \textbf{b.} azimuth ($\phi$) at 28GHz. \textbf{c.} Measured constellations and EVMs at different scan angles after transmission over 35m fiber and 0.45m wireless link using 400Mbaud (2Gbit/s) 32QAM modulation.} \label{fig:4}
\end{center}
\end{figure}
%%%%%%%%%%%%%%%%%%%%%%%%%%%%%

\section*{Discussion and conclusion}

Hybrid metasurfaces incorporating optic, photonic and electronic techniques have the potential to address the requirements of next-generation wireless communications, compute systems, advanced imagers, and sensors to name a few. At the same time, these technologies offer the advantages of simpler architectural designs. Nevertheless, for real-world applications and effective deployment, several avenues should be investigated.\par
In the demonstrated proof-of-concept reconfigurable non-linear active metasurface, microlenses are used to enhance the optical coupling from free-space to the chips. However, given that the beam-waist is expanded to a diameter large enough to illuminate all four chips, only a fraction of the light is collected by the aperture of the lens, focused on the grating couplers, and eventually photodetected. Alternatively, beam splitting using diffractive optical elements \cite{katz2018using,wang2020toward, wu2017design} can be used. Such splitters can generate uniform two-dimensional spot arrays with specified separations and high efficiencies (>80\%)\cite{Holo2020}. The EPIC chips could then be positioned on the location of these spots to couple in the light, ensuring minimal optical power is incident on regions without any chips. By employing such optical elements, coupling efficiency can be further improved and a higher EIRP could be achieved using lower optical power leading to reductions in system power consumption. Furthermore, enhancing optical coupling efficiency would permit operation at lower optical powers, thereby avoiding nonlinear losses in the fiber (which can limit the power reaching the metasurface) and hence facilitate transmission over extended distances. A potential challenge associated with the transmission of mm-wave carrier frequencies over long distances of fiber may be power fading induced by chromatic dispersion\cite{gliese1996chromatic}, but this issue can be mitigated using techniques such single sideband modulation or optical filtering\cite{martin201728,tong2019integrated}.\par
In the context of communication links and distribution of signals to the metasurface, the majority of data transmission takes place over fiber and mm-wave wireless link. However, it is worth mentioning that it is also possible to further extend the transmission distance by positioning the collimator further away from the metasurface, thereby increasing the free-space optical path between them. As detailed in the Methods section, given the effective focal length of the collimator, the output free-space beam can remain collimated for approximately 12.6m (compared to the 8 cm distance from the metasurface to the collimator in the presented measurements). Note that though for practical applications, this scheme becomes more prone to atmospheric variation and or obstructions which can lead to, for example, increased side-lobe levels if the optical power is unequally distributed to the different elements. In practice, amplitude variations can be compensated to a certain extent by monitoring the photocurrents of the individual photodetectors and adjusting the on-chip variable gain amplifiers which can provide 2.5dB gain control (Extended Data Fig. 4d) while introducing a phase change of less than 4$^\circ$ (Extended Data Fig. 4e). Note that for 5dB gain control, the phase variation increases to 10$^\circ$.\par
The prototype metasurface is comprised of electronic-photonic chips that were implemented using 90nm CMOS technology. The more advanced node of this technology, GlobalFoundries 45CLO process\cite{rakowski202045nm}, offers higher bandwidth photodetectors (50GHz) and 45nm CMOS transistors with higher transit frequencies paving the way for operation at higher mm-wave frequencies and ultimately enabling higher data-rates and lower power consumption. Having said that, even with the presented system, we can expect to achieve higher-data rates by using higher modulation orders and or larger modulation bandwidths given that the bandwidth of the system is approximately 1.5GHz as shown by the measurements of Extended Data Fig. 4c. However, as previously explained and elaborated in the Methods section, the performance and data rate were largely constrained by the SNR and available equipment and can be expected to increase by improving the optical coupling scheme.\par
As an alternative to waveguide coupled photodetectors, optimizing the photodetectors for vertical illumination\cite{kim2013high} can circumvent the coupling losses associated with the grating coupler and thus improve system efficiency. Though a potential challenge with this approach may be the trade-off between the speed and responsivity of the device, along with the limited access to the doping profiles and dielectric stackup of the foundry process required for optimization. Another approach may be to employ photodetectors with high bandwidths and high quantum efficiencies which are heterogeneously integrated and bonded, or co-packaged with CMOS (or BiCMOS) integrated circuitry. This approach can increase the flexibility in optimizing the electronic and photonic platforms independently and may also potentially relax the sensitivity to optical alignment as the photodetector aperture can be larger when compared to a grating coupler but comes at the expense of introducing a more intricate packaging process.\par
Finally, to realize a metasurface with a larger number of elements, a tiling technique may be employed. In this regard, the unit-cell EPICs lend themselves conveniently as they can be tiled without restriction. The chips could then be grouped into sub-modules with each sub-module having a dedicated diffractive optical element or lens array. Light from the laser could then be split using fiber optic couplers (splitters) and distributed to the various submodules similar to arrangements previously explored for optically synchronized phased arrays\cite{gal2022optically}.\par
In summary, we have demonstrated, for the first, a reconfigurable non-linear active metasurface capable of radiating a data-carrying mm-wave signal when illuminated by an optical wave that has been modulated with the mm-wave signal. The metasurface is comprised of electronic-photonic chips, each integrating a photonic receiver and mm-wave transmitter, tiled on a low-cost printed circuit board with an antenna array and is designed for operation around 193 THz at its input and 28GHz at its output. With the metasurface, we showcase beam steering with a scanning range of approximately 60$^\circ$ in elevation and azimuth. Additionally, a communication link is demonstrated where data is transmitted over optical fiber, coupled to the metasurface through free-space and wirelessly radiated at 28GHz in various directions, achieving data rates as high as 2Gb/s.

\section*{Methods}

\subsection*{Detection of the mm-wave signal} In Fig. 2a, the laser emitting power $P_o$ at angular frequency $\omega_o$, is intensity-modulated by a mm-wave signal of angular frequency $\omega_{mm}$ using a Mach-Zehnder modulator (MZM). The electric field at the output of the MZM can be expressed as 

\begin{equation}
E_{MZM} = \sqrt{P_o}e^{j\omega_o t}e^{j\theta /2}e^{j\beta A\cos{(\omega_{mm}t)}/2} \cos \left( \frac{\beta A}{2} \cos ( \omega_{mm}t) + \frac{\theta}{2} \right)
,\label{eq1}
\end{equation} 
\noindent where $\beta = \pi / V_\pi $, $V_\pi$, $A$, $\theta$ are the modulation index, the modulator voltage required to induce a $\pi$ phase shift, the amplitude of the driving mm-wave signal, and the phase difference between the output arms of the MZM, respectively. The light at the output of the MZM is then expanded and collimated to a spot-size $A_{spot}$ large enough to illuminate the EPIC chips that have been tiled on the top side of the printed circuit board. The modulated optical wave couples into these chips through grating couplers and is guided to photodetectors with silicon nanophotonic waveguides. The photocurrent generated at the output of a photodetector $I_{PD}$ can be calculated as $I_{PD}=EE^*$  where $E$ is the electric field coupled to the chip. By using the Jacobi-Anger expansions and setting $\theta = \pi / 2$, the generated photocurrent, in the absence of a microlens array, can be approximated as 

\begin{equation}
I_{PD} \approx I_{DC} + \alpha \frac{A_{GC}}{A_{spot}}P_o R J_1(\beta A)\cos(\omega_{mm}t) 
,\label{eq2}
\end{equation} 

\noindent where $J_1(.)$, $I_{DC}$, $A_{GC}$, and $\alpha$ are the 1st order Bessel function of the first kind, the DC component of the photocurrent, the area, and the coupling factor of the grating coupler, respectively. Note that in practice, the light exiting the collimator has a Gaussian distribution; however, for simplicity, we have assumed that the output power is uniformly distributed within the spot size. As evident from eq (\ref{eq1}), the output photocurrent has a mm-wave component which is amplified by the transmitter chain.

\subsection*{Lens-enhanced optical coupling}
Considering a grating coupler area of $A_{GC}$, and a collimated beam spot size of $A_{spot}$, the amount of optical power coupled from free space into a grating coupler can be calculated as $\alpha A_{GC} / A_{spot}$ which in our design results in around 64.5dB coupling loss. Since there are four grating couplers per chip, the effective coupling loss per chip is around 58.5dB. When a microlens is used, as illustrated in Fig 2a, the lens collects and focuses a larger amount of power, but only on one of the grating couplers of a chip. In this case, for a lens area of $A_{lens}$, and given that the focused spot size is smaller than $A_{GC}$, the optical power coupled to the chip significantly increases. In this case, the $A_{GC} / A_{spot}$  factor in eq (\ref{eq1}) is modified to $A_{lens} / A_{spot}$. Through the manual alignment of a lens, the coupling loss per chip was lowered to around 39.5dB.

\subsection*{Microlens alignment} 
Off-the-shelf spherical microlenses (Edmund Optics \#21-178), which have been fabricated on a silicon substrate, were used to focus the optical wave on the grating couplers of the EPICs. The microlenses, which have a diameter of 0.69mm and a radius of curvature of 3.2mm, were actively aligned with the chips. To achieve this, a lens was loosely attached to a fixture placed on a 3-axis precision stage (Thorlabs 3-axis NanoMax). A collimated laser light was incident on the chips and the DC currents of the photodiodes were monitored. Using the precision stage, the lens was positioned and aligned above a grating coupler (close to the working distance) such that the DC current of the corresponding photodiode was maximized. To set the position of the lens, UV-cured glue was applied to fix the lens on a microscope slide that had been placed above the chips (secured to the PCB). Once the glue was cured, the loosely attached fixture holding the microlens and connected to the 3-axis stage was removed. Note that in this proof-of-concept demonstration, the alignment and gluing were carried out manually and as such the best precision was not necessarily attained; however, as shown by the results of Extended Data Fig. 4b, the EIRP increases by around 38dB corresponding to a 19dB improvement in optical coupling. 

\subsection*{Design of the electronic-photonic integrated circuit}
The electronic-photonic integrated chip was fabricated on the GlobalFoundries 90nm CMOS silicon photonic process. In this process, the NFET has a ft/fmax of 148/224GHz and a nominal supply voltage of 1.2V\cite{giewont2019300}. The circuit topology of the various blocks in the photonic receiver and transmitter chain are shown in Extended Data Fig. 1. Differential structures with cascode topology are generally preferred to ensure circuit stability. The mm-wave phase shifter is composed of a one stage polyphase filter\cite{kaukovuori2008analysis} generating quadrature signals and two Gilbert cells forming the vector modulators\cite{koh20070}. The polyphase filter has a relatively large insertion loss of around 15dB whereas the active vector modulator acts as a buffer. The pre-driver and power amplifier are implemented as pseudo-differential cascode stages with cross-connected neutralization capacitors. The transformers used for inter-stage matching were designed and simulated using Ansys HFSS 3D electromagnetic simulator. To reduce the inductive effects of the signal bond wires, the chips were backgrinded from 750$\mu$m to 250$\mu$m. The on-chip integrated photonic components include grating couplers, optimized for transverse-electric (TE) polarization with a 28\% (5.5dB) coupling efficiency, strip waveguides with a loss of around 1.5dB/cm, and SiGe p-i-n photodetectors with a responsivity around 1A/W and an electro-optic bandwidth of approximately 40GHz.

\subsection*{Stand-alone characterization of the electronic-photonic chip}
The electronic-photonic chips were designed to operate with optical input and electrical output. However, to accurately characterize the performance of the various blocks, an electrical differential test input that bypasses the photodetectors and is connected to the input of the first stage amplifier was included on chip. The s-parameter results shown in Fig. 2a are measured using a differential test setup. A Vector Network Analyzer (VNA) and high-frequency differential probes (calibrated to the probe tips) were used to excite the chip input and measure the signal at its output. Note that the output of the chip has been designed to drive the wire-bonded differential patch antenna and not a 100$\Omega$ differential load. To characterize the vector modulator phase shifter, the phase and gain responses of the chip were recorded using the VNA. Extended Data Fig. 3a shows the measured insertion phases when the on-chip phase shifter was programmed for 36 different phase states at 10$^\circ$ spacings. Extended Data Fig. 3b shows the corresponding gain change for the different phase states. From these data, the average RMS phase and amplitude errors were calculated, which are shown in Extended Data Fig. 3c. At 28GHz, the measured RMS phase error is around 4$^\circ$ and the average RMS amplitude error is around 7\%. To further reduce the phase and gain errors, the phase and gain responses were recorded using the VNA, and a gradient descent algorithm was implemented to find the appropriate VM and VGA settings that resulted in the desired phase shifts and amplitudes. Figure 2e shows the results of the calibration at 28GHz, where the phase error has been successfully reduced to 1.8$^\circ$ and the average RMS amplitude error has been lowered to 3.5\%. Note that these corresponding settings were then stored in a look-up table and used for the demonstrated beam scanning measurements.

\subsection*{Effect of test input traces on trans-impedance gain}
The differential electrical test input traces enable probing for electrical-in electrical-out measurements; however, for optical measurements, they result in a lower trans-impedance gain (at the input) and as such were designed to be laser trimmed. To demonstrate this, light from a laser, that had been intensity modulated at $\omega_{mm}$, was coupled into one of the grating couplers of the EPIC using an optical fiber probe. The output of the chip was electrically probed and the output power at $\omega_{mm}$ was then measured both before and after laser trimming the test input traces. Extended Data Fig. 4a shows the effect of the test input trace on the system where almost 11dB gain improvement is observed when the test input traces are laser-trimmed.

\subsection*{Metasurface antenna array design}
The 2$\times$2 patch antenna array of the metasurface was designed and simulated using Ansys HFSS 3D electromagnetic simulator. Extended Fig. 2 shows the dimensions of the patch antenna and stackup of the low-cost 4-layer FR4 printed circuit board. The patch is implemented on the bottom layer (L4) and the antenna feedline, which is wire-bonded to the output of the chip and connects to the patch through differential vias, is implemented on the top layer (L1) with the two middle layers (L2 and L3) used as ground planes. The differential impedance of the antenna is around 75$\Omega$. The array achieves a simulated gain of around 9dB.

\subsection*{Beamformation and radiation pattern measurements}
Light from a CW laser (Optilab DFB-40R-20-P) operating near 1550nm had its polarization set using a polarization controller and was intensity modulated with a 28GHz (Anritsu MG3697C) mm-wave signal using a 40GHz intensity modulator (Thorlabs LN05S-FC). The intensity-modulated light was amplified using an EDFA (Nuphoton Technologies EDFA-CW-C0-RS-33-36-FCA) and collimated using a collimator (Thorlabs F810APC-1550) creating an optical beam with a waist diameter of around 7mm which was used to illuminate the metasurface. To receive the radiated 28GHz mm-wave beam, a standard WR-28 horn antenna, with a nominal gain of 10dB, was placed 45cm away in the far-field of the antenna array. The horn antenna was mounted on a custom-made robotic arm that could scan the surface of a half-sphere in order to record the received power at different points in elevation and azimuth. The signal was then amplified by a low-noise amplifier (Narda-MITEQ AMF-6F-26003300-50-10P) – which has a gain of around 40dB and noise figure of 5dB – and monitored on a signal analyzer (Agilent PXA N9030A). Using the calibrated phase shifter settings that were recorded during the characterization of the electronic-photonic chip, the metasurface was configured to point at different elevation (azimuth) angles. To measure the radiation pattern for each beam direction, the robotic arm was scanned from -60$^\circ$ to +60$^\circ$ in elevation (azimuth) in 2$^\circ$ steps and the received power was measured at each point.

\subsection*{Data communication}
Matlab was used to create a 400Mbaud 32QAM (2Gbit/s) waveform, filtered with a root-raised cosine pulse shaping filter with a roll-off factor of 0.35, at an intermediate frequency of 7GHz (IF1). The waveform was uploaded to the memory of a Micram DAC10002 running at 96GS/s and serving as the AWG. The output of the AWG (QAM signal at 7GHz) was up-converted to 28GHz using a waveguide mixer (Fairview FMMU1005) and up-converting sinusoidal LO1 at 35GHz (generated using Anritsu MG3696A). The choice of LO1 frequency was due to the frequency range of the available mixer, but an LO1 at 21GHz could also be used with an appropriate mixer. A bandpass filter (Pasternack PE8747) was used to filter out the image signal at 42GHz. A 25dBm linear power amplifier (Toshiba BA2076B), operating in backoff to minimize EVM degradation, was used to amplify the filtered signal and drive a 40GHz intensity modulator (Thorlabs LN05S-FC). Light from a CW laser (Optilab DFB-40R-20-P) operating near 1550nm was intensity modulated after having its polarization adjusted to maximize modulation efficiency and amplified using an EDFA (Nuphoton Technologies EDFA-CW-C0-RS-33-36-FCA). The polarization of the amplified light was adjusted once more to maximize coupling into the grating couplers optimized for TE-polarization. The modulated optical signal was transmitted through a 35m single-mode fiber and connected to a collimator (Thorlabs F810APC-1550) to generate a free-space collimated beam. \par
At the receiver, the 28GHz mm-wave beam was received by a linearly polarized WR-28 horn antenna, with a gain of around 10dB, and amplified using a low-noise amplifier (Narda-MITEQ AMF-6F-26003300-50-10P). The signal was down-converted to 1.5GHz (IF2) using an external mixer (Marki M20240LP) and second sinewave LO2 at 26.5 GHz (generated using Anritsu MG3697C). The down-converted signal was then amplified (using Mini-Circuits ZX60-6013E-S+) and monitored on a real-time oscilloscope (Keysight EXR258A). Carrier-recovery, digital demodulation, and signal analysis were performed in real-time on the oscilloscope using the Keysight Pathwave Vector Signal Analysis software.

\subsection*{EIRP and wireless transmission distance calculations}
The equivalent isotropic radiated power (EIRP) was calculated using the equation 
\begin{equation}
EIRP = P_R - G_R + L_S
,\label{eq3}
\end{equation} 
\noindent where $P_R$, $G_R$, and $L_S$ are the received power (in dBm), the gain of the receiver, and the free space propagation loss (both expressed in dB) respectively. $P_R$ was measured using a calibrated signal analyzer. $G_R$ consisted of the gains from the horn antenna, the low-noise amplifier, and the losses incurred through the cables. Measurements were done at a distance of 45cm corresponding to $L_S$ of 54.5dB at 28GHz.\par
In the presented work, an EIRP 10dBm was calculated when operating with 200mW optical power. To estimate the maximum wireless transmission distance, the minimum required signal at the input of a wireless receiver can be estimated as \cite{pozar2000microwave}

\begin{equation}
P_{RS} = 10\log_{10}\left(\frac{kT}{1mW}\right)+NF+10\log_{10}\left(BW\right)+SNR - G_{ant},
\label{eq4}
\end{equation} 

\noindent where $k$, $T$, $NF$, $BW$, $SNR$, and $G_{ant}$ are the Boltzmann constant, temperature, noise figure of the receiver, bandwidth of the signal, SNR required by the receiver and antenna gain respectively. For a 32QAM signal, to achieve a bit-error rate of $10^{-6}$, a SNR of around 23dB is required\cite{goldsmith2005wireless}. Given a temperature of 300K, a bandwidth of 400(1+0.35)MHz, a noise figure of 5dB, an antenna gain of 10dB, the minimum required signal level is around -69dBm. Hence, with an EIRP of 10dBm, the path loss that can be tolerated is around 79dB which corresponds to a wireless transmission distance of 9m when operating at 28GHz.

\subsection*{SNR calculations} 
In the communication link presented in Fig. 3b, the AWG has an average output power ($P_{AWG}$) of -13dBm and a signal-to-noise ratio ($SNR_{AWG}$) of around 32dB (ENOB of 5 bits) when generating a 400Mbaud QAM32 waveform at a frequency of 7GHz. The mixer (Fairview FMMU1005) has a conversion loss of 6dB and the filter (Pasternack PE8747) has an insertion loss of 2dB at 28GHz. The amplifier driving the intensity modulator has a gain of 35dB and thus the power delivered to the intensity modulator is around 14dBm (25mW). Note that the SNR is not degraded by the noise figure of the cascaded mixer, filter and driver amplifier as the output noise power spectral density (PSD) of the AWG ($S_{n,AWG}$) is much higher than the thermal noise floor (i.e. -174dBm/Hz). That is 
\begin{equation}
\begin{aligned}
S_{n,AWG} &= P_{AWG} - SNR_{AWG} - 10\log_{10}\left(BW\right) \\
          &= 132\text{dBm/Hz,}
\end{aligned}
\label{eq5}
\end{equation} 
\noindent where $BW=400(1+0.35)$MHz and 0.35 is the roll-off factor of pulse-shaping filter. \par
Following a similar derivation to eq.(\ref{eq1}) with the difference that the driving signal $A\cos(\omega_{mm})$ is replaced with $v(t)$, the electrical field at the output of the MZM can be described as 
\begin{equation}
E_{MZM} = \sqrt{P_{o}}e^{j\omega_{o}t}e^{j\theta/2}e^{j\beta v(t)/2} \cos \left(\frac{v(t)}{2}+\frac{\theta}{2}\right).
\label{eq6}
\end{equation} 
\noindent By setting $\theta=\pi/2$ and assuming small signal modulation( i.e. $\sin\left(\beta v(t) \right) \approx \beta v(t)$ ), the generated photocurrent within each chip can be approximated as
\begin{equation}
I_{PD} \approx \frac{\alpha_e P_o R}{2}\left( 1+\beta v(t) \right)
\label{eq7}
\end{equation} 
\noindent where $ \alpha_{e}$ is the effective coupling factor and includes the effect of the microlens and loss of the grating coupler. Since the signal and AWG noise experience the same link gain, the variance of the current fluctuations due to the AWG at the output of the photodetector can be expressed as 
\begin{equation}
 \overline{i_{n,AWG}^2} = \overline{i_{sig}^2}/SNR_{AWG},
\label{eq8}
\end{equation}
\noindent where $\overline{i_{sig}^2} = \frac{1}{4}\left( \alpha_e P_o R \beta\right)^2 \overline{v(t)^2}$. \\
\noindent Considering the various sources contributing to noise, the total noise current in at the output of a photodetector can be written as
\begin{equation}
 i_n = i_{n,T}+i_{n,shot}+i_{n,RIN}+i_{n,sp-sp}+i_{n,sig-sp}+i_{n,tx}+i_{n,AWG}
\label{eq9}
\end{equation}
\noindent where $i_{n,T}, i_{n,shot}, i_{n,RIN}, i_{n,sp-sp}, i_{n,sig-sp},$ and $i_{n,tx}$ are current fluctuations due to the thermal noise, shot noise of the photodetector, the relative intensity noise (RIN) of the laser, beating of the optical field associated with the amplified spontaneous emission of the EDFA with itself, beating of the optical field associated with the amplified spontaneous emission of the EDFA with the signal, and the input referred current noise of the electronics chain respectively \cite{agrawal2012fiber}. To find the dominant terms, the variance of the above terms (or noise power) can be calculated and are given as
\begin{equation}
\begin{aligned}
&\overline{i_{n,T}^2} = 4kT \Delta f/R_L = 4.97 \times 10^{-13}\\
&\overline{i_{n,shot}^2} = q\alpha_eRP_o \Delta f = 5.38 \times 10^{-15}\\
&\overline{i_{n,RIN}^2} = (\alpha_eRP_o)^2 RIN \Delta f = 7.55 \times 10^{-15}\\
&\overline{i_{n,sp-sp}^2} = (\alpha_e q \eta G_{EDFA} NF_{EDFA})^2 \Delta \nu_{EDFA} \Delta f = 2.29 \times 10^{-19} \\
&\overline{i_{n,sp-sig}^2} = 2(\alpha_e q \eta)^2 G_{EDFA} NF_{EDFA} P_o \Delta f / h \nu = 4.30 \times 10^{-16} \\
&\overline{i_{n,AWG}^2} = 0.25(\alpha_e R P_o \beta)^2 \overline{v(t)^2} / SNR_{AWG} = 2.72 \times 10^{-14}\\
&\overline{i_{n,tx}^2} = 2.16 \times 10^{-13} \text{(simulated)}\\
\end{aligned}
\label{eq10}
\end{equation} 
where $k, T, \Delta f, RIN, q, G_{EDFA}, NF_{EDFA}, \Delta \nu_{EDFA}, h, \nu, \eta=R h \nu / q$ are the Boltzmann constant, temperature, bandwidth of the TIA, RIN of the laser, elementary charge, EDFA gain, EDFA noise figure, EDFA bandwidth, Planck constant, the optical field frequency and quantum efficiency of the photodetector respectively. Note in the above calculations, we have used the measurement setup values of $P_o=\text{200mW}$ (output of EDFA), $R=1\text{A/W}$, $V_{\pi}=\text{6V}$, $\alpha_{e}=\text{-39.5dB}$, $R_L=\text{50}\Omega$, $RIN=\text{-140dBc/Hz}$, $G_{EDFA}=\text{100}$, $NF_{EDFA}=\text{6.5dB}$, $\Delta f=\text{1.5GHz}$, $\Delta \nu_{EDFA}=\text{3.75THz ($\sim$ 30nm)}$, $\nu=\text{193THz}$ and the fact that $\overline{v(t)^2}/50=\text{25mW}$ (intensity modulator has an impedance of 50$\Omega$).

Considering only the effects of the thermal noise, AWG noise and input referred noise of the electronics, the SNR at the output of the photodetector can be approximated as
\begin{equation}
 SNR = \frac{\overline{i_{sig}^2}}{\overline{i_{n,T}^2} + \overline{i_{n,AWG}^2} + \overline{i_{n,tx}^2} } \approx 17.5\text{dB}.
 \label{eq11}
\end{equation}
\noindent The metasurface increases the SNR by $10\log_{10}(4)=6\text{dB}$ to 23.5dB since the noise of the input of each channel is uncorrelated whereas the signals are correlated. At the receiver, an EVM of around 6\% is measured for a QAM32 waveform which corresponds to an SNR of 20.5dB. As such, the wireless link and receiver chain incur an SNR penalty of around 3dB. From the above calculations, it is evident that given the detected optical power, the system operates in the thermal noise limited regime and the SNR of the whole link degrades from 32dB (SNR at output of AWG) to 20.5dB. Therefore, by improving the optical coupling efficiency and using better equipment (i.e. higher $SNR_{AWG}$), it is expected that the signal power can be increased while maintaining a constant noise floor until the shot-noise limited regime is reached. This in turn will increase the SNR and enable higher order modulation formats and data-rates to be achieved. Note in general, with higher optical powers (better coupling efficiency), the RIN of the laser will also contribute additional noise.

\subsection*{Free-space optical path}
The collimator (Thorlabs F810APC-1550) used to illuminate the metasurface has an effective focal length (EFL) of 37.13mm. The maximum distance d for which the output beam remains collimated can be approximated as
\begin{equation}
d = EFL + \frac{2(EFL)^2 \lambda}{\pi (MFD)^2},
\label{eq12}
\end{equation}
\noindent where $MFD$ and $\lambda$ are the mode-field diameter of the single mode fiber and the operating wavelength respectively. For $MFD=\text{10.4}\mu m$ and $\lambda = \text{1550nm}$, distance $d$ is approximately equal to 12.6m. In the measurements, the distance between the collimator and metasurface was approximately 8cm which could theoretically be extended to 12.6m to, for example, facilitate transmission over a longer free-space optical path.

\section*{Data availability}
The data that support the findings of this study are available from the corresponding author upon request.

% Pouria - overwrite settings
\def\figurename{Extended Data Fig.}
\setcounter{figure}{0}

\clearpage
\newpage
%%%%%%%%%%%EXTENDED FIGURE%%%%%%%%%%%%%%%%%%
\begin{figure}[htb]
\begin{center}
\includegraphics[width=176mm]{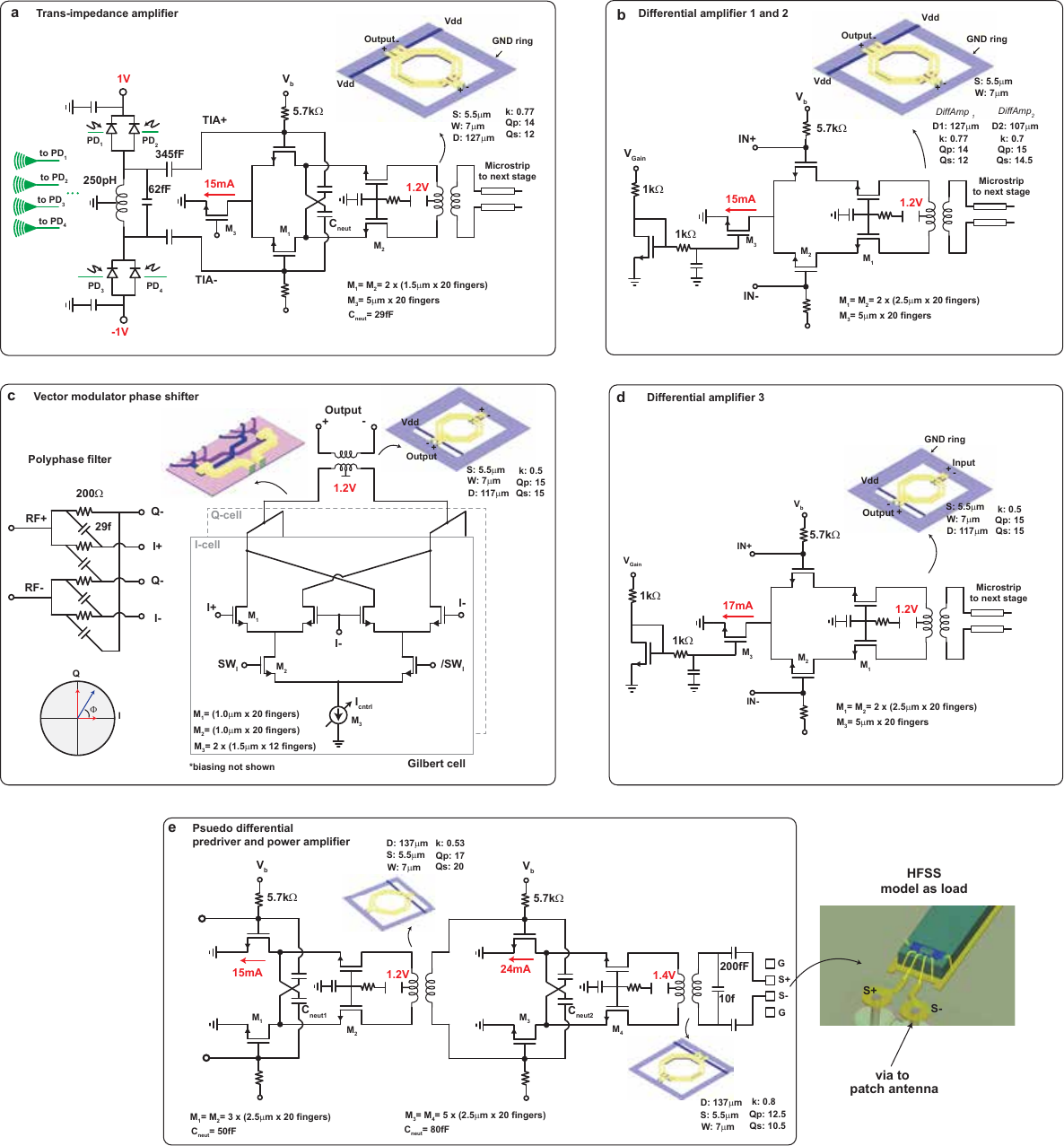}
\caption{\textbf{Circuit topology of various blocks.} \textbf{a.} Schematic of the trans-impedance amplifier implemented using a differential cascode with cross-connected capacitors for gate-drain neutralization and a symmetric inductor with a grounded center tap that resonates with the capacitances of the photodetectors and input transistors at 28GHz. \textbf{b.} Schematic of the variable gain differential cascode amplifiers. \textbf{c.} Schematic of the active vector modulator phase shifter. It is composed of a one stage polyphase filter generating I/Q signals and two Gilbert cells forming the vector modulators. Vector synthesis is achieved with tail current controls and quadrant selection through the switches ($SW_I$ and $SW_Q$). \textbf{d.} Schematic of the differential cascode amplifier between the phase shifter and pre-driver. \textbf{e.} Schematic of the pseudo-differential predriver and power amplifier employing cross-connected capacitors for gate-drain neutralization.} \label{fig:e1}
\end{center}
\end{figure}
%%%%%%%%%%%%%%%%%%%%%%%%%%%%%

\clearpage
\newpage
%%%%%%%%%%%EXTENDED FIGURE%%%%%%%%%%%%%%%%%%
\begin{figure}[htb]
\begin{center}
\includegraphics{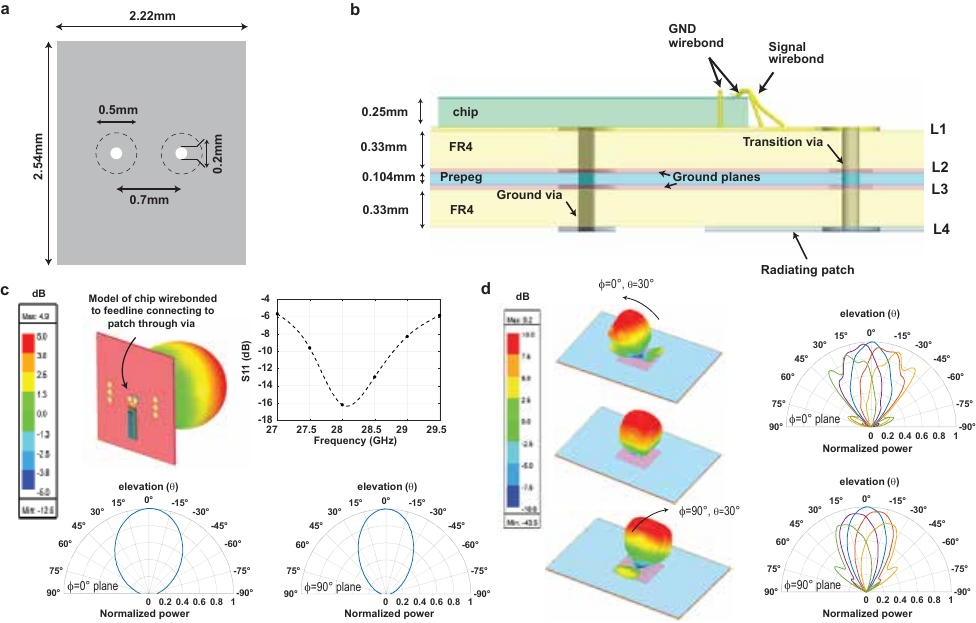}
\caption{\textbf{PCB and antenna array design.} \textbf{a.} Dimensions of the 28GHz differential microstrip patch antenna. \textbf{b.} Stackup of the printed circuit board. The feedlines of the antennas, that are wire bonded to outputs of the electronic-photonic chips, are implemented on layer 1 (L1) and connect to the radiating patches, implemented on layer 4 (L4), using transition vias with layer 2 (L2) and layer 3 (L3) used as ground planes. The DC power lines, and digital control lines used to adjust the gain and phase settings of the chips are implemented on L1. \textbf{c.} Ansys HFSS simulation of a single microstrip patch indicating a gain of around 4.4dB and differential impedance of around 75$\Omega$. \textbf{d.} Ansys HFSS simulation of 2$\times$2 patch antenna array (surrounded by dummy patches) with an array gain of around 9dB (broadside configuration) and beam-steering demonstration in two planes.} \label{fig:e2}
\end{center}
\end{figure}
%%%%%%%%%%%%%%%%%%%%%%%%%%%%%

\clearpage
\newpage
%%%%%%%%%%%EXTENDED FIGURE%%%%%%%%%%%%%%%%%%
\begin{figure}[htb]
\begin{center}
\includegraphics[width=176mm]{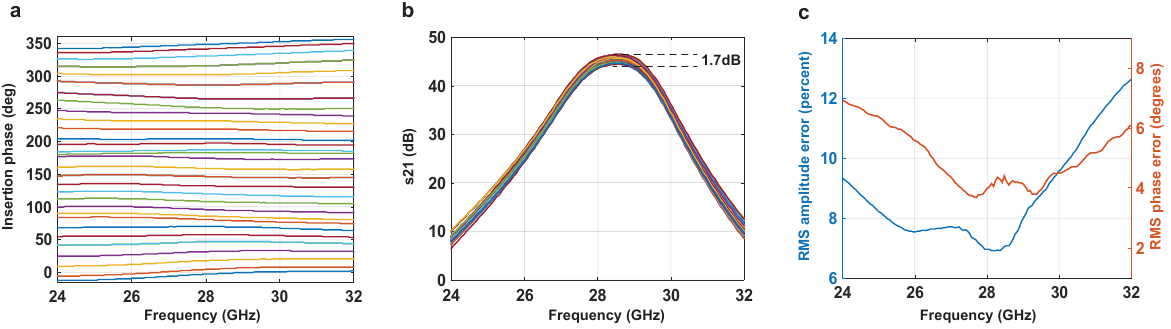}
\caption{\textbf{Vector modulator phase shifter performance before calibration.} \textbf{a.} Measured insertion phases for 36 different phase states at 10$^\circ$ spacings. \textbf{b.} Corresponding gain change over the various phase states. \textbf{c.} Calculated RMS amplitude and phase errors.} \label{fig:e3}
\end{center}
\end{figure}
%%%%%%%%%%%%%%%%%%%%%%%%%%%%%

\clearpage
\newpage
%%%%%%%%%%%EXTENDED FIGURE%%%%%%%%%%%%%%%%%%
\begin{figure}[htb]
\begin{center}
\includegraphics[width=176mm]{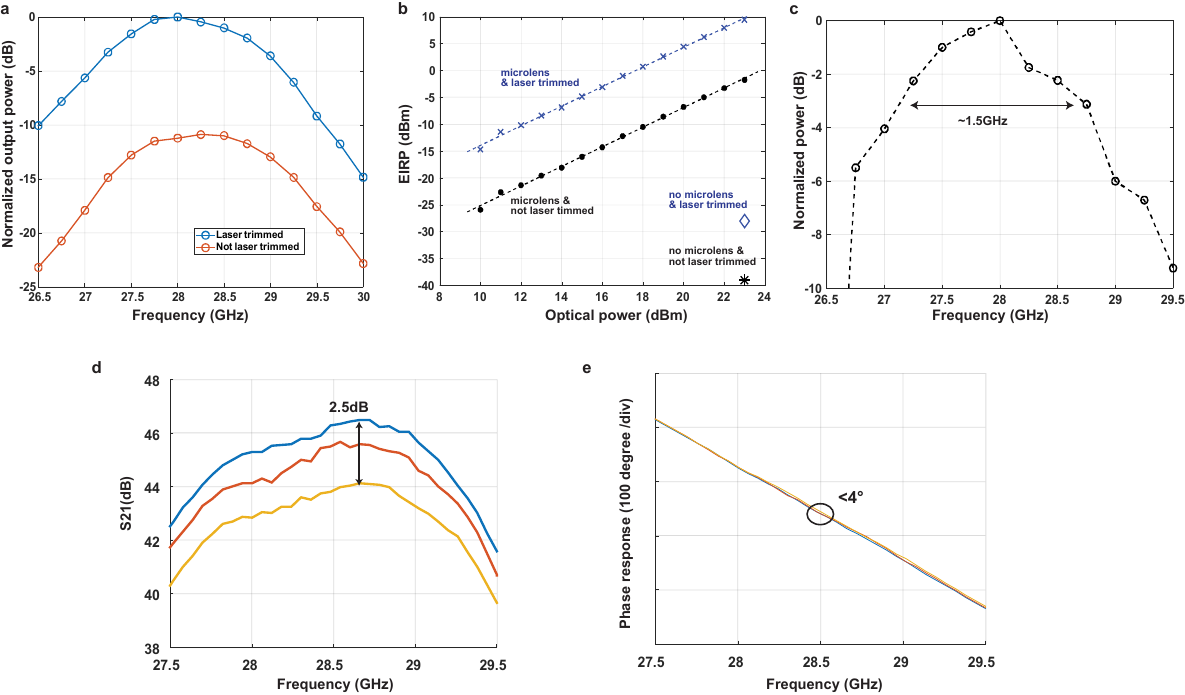}
\caption{\textbf{Additional measurements.} \textbf{a.} Effect of test input traces on the gain of the EPIC. For the same optical power, the RF power at the output of the chip increases by around 11dB when the test input traces are laser trimmed. \textbf{b.} EIRP of the metasurface vs optical power of the laser considering the effects of the microlens and test input RF traces. \textbf{c.} Bandwidth of the metasurface. Here, the frequency of the mm-wave signal driving the intensity modulator is swept and the power received by the horn antenna is recorded.} \label{fig:e4}
\end{center}
\end{figure}
%%%%%%%%%%%%%%%%%%%%%%%%%%%%%

\newpage
\clearpage
\bibliography{main}

\section*{Author contributions statement}

P.S. and F.A. conceived the project idea. P.S. designed, simulated, and laid out the chip and PCB. P.S. conducted measurements. F.A. directed and supervised the project. P.S. and F.A. wrote the manuscript.

\section*{Competing interests}
The authors declare no competing interests.

\section*{Materials \& correspondence}
Correspondence and request for materials should be addressed to Firooz Aflatouni.

\end{document}